\documentclass[%
 reprint,
superscriptaddress,
 nofootinbib,
 nobibnotes,
 amsmath,amssymb,
 aps,
 prl,
 floatfix,
]{revtex4-1}

\usepackage{graphicx}
\usepackage{dcolumn}
\usepackage{bm}
\usepackage[T1]{fontenc}
\usepackage[utf8]{inputenc}
\usepackage{subfigure}
\usepackage{hyperref}
\usepackage{xcolor}
\usepackage[english]{babel}
\usepackage{soul} 

\usepackage{xspace}

\newcommand{\ii}{\ensuremath{\mathrm{i}}\xspace}

\newcommand{\e}[1]{\ensuremath{\mathrm{e}^{#1}}}

\AtEndDocument{%
    \newwrite\bibnotes
    \def\bibnotesext{Notes.bib}
    \immediate\openout\bibnotes=\jobname\bibnotesext
    \immediate\write\bibnotes{@CONTROL{REVTEX41Control}}
    \immediate\write\bibnotes{@CONTROL{%
    apsrev41Control,author="08",editor="1",pages="1",title="0",year="1"}}
     \if@filesw
     \immediate\write\@auxout{\string\citation{apsrev41Control}}%
    \fi
}%

\begin{document}


\title{
Topological Effects of a Vorticity Filament on the Coherent Backscattering Cone
}

\author{Geoffroy J. \surname{Aubry}}%
\email{geoffroy.aubry@unifr.ch}
\altaffiliation[Now at: ]{%
 Département de Physique, Université de Fribourg, 1700 Fribourg, Switzerland
}%
\affiliation{%
 Fachbereich Physik, Universität Konstanz, 78457 Konstanz, Germany
}
\author{Philippe \surname{Roux}}
\email{philippe.roux@univ-grenoble-alpes.fr}
\affiliation{%
  Université Grenoble Alpes, Université Savoie Mont Blanc, CNRS, IRD, IFSTTAR, ISTerre, 38000 Grenoble, France 
}%

\date{\today}

\begin{abstract}
In this Letter, we report on the effects of a vorticity filament on the coherent backscattering cone. Using ultrasonic waves in a strongly reverberating cavity, we experimentally show that the discrete number of loops of acoustic paths around a pointlike vortex located at the center of the cavity drives the cancellation and the potential rebirth of the coherent backscattering enhancement. The vorticity filament behaves, then, as a topological anomaly for wave propagation that provides some new insight between reciprocity and weak localization.
\end{abstract}

\maketitle


Coherent backscattering enhancement of waves by a random medium provides convincing evidence of interference effects despite disorder and multiple scattering.
The coherent backscattering cone (CBC) is manifested as a cusp in the angular distribution of the backscattered intensity~\cite{Akkermans1986}.
This universal phenomenon has been observed experimentally at different spatial scales in optics~\cite{Wolf1985,*Albada1985}, acoustics~\cite{Bayer1993}, and seismology~\cite{Larose2004}, and with elastic waves~\cite{Rosny2000} and cold atoms~\cite{Jendrzejewski2012a}.  

The CBC is considered to be a sign of weak localization of waves that propagate in a disordered medium.
The weak localization originates from constructive interferences between multiple scattering paths and their reciprocal counterparts that follow the same sequence of scatterers in reverse order.
Indeed, only reciprocity breaking in the propagation medium can alter the constructive interferences and destroy the enhancement~\cite{Golubentsev1984}.
Suppression of the CBC by Faraday rotation of light in a multiple scattering medium provided the first experimental evidence of the disappearance of weak localization~\cite{Erbacher1993,*Schertel2017}. 
Evidence of CBC destruction using acoustic waves in a rotational flow was also reported~\cite{Tourin1997}.
The relation between reciprocity and coherent backscattering enhancement was recently revisited in the context of propagation of cold atoms in optical speckles~\cite{Mueller2015} or of light in optical fibers~\cite{Bromberg2016}.

In this Letter, we go one step beyond the CBC destruction for acoustic waves by  observing a coherent backscattering dip and by predicting the rebirth of the cone. Compared to the reciprocity-breaking phase shifts induced by solid rotation, which depends on the path lengths distribution in the cavity, the topological anomaly created by a vorticity filament quantizes wave transport properties: the \emph{only} physical parameter that drives the CBC cancellation and its potential rebirth is the \emph{integer number} of loops of the conjugated paths around this topological singularity. Note also that the interplay between topology and hydrodynamics is more general, e.g., a recent study of the topological nature of some geophysical flows like the El Ni\~no phenomenon~\cite{Delplace2017}.

\begin{figure}
\includegraphics{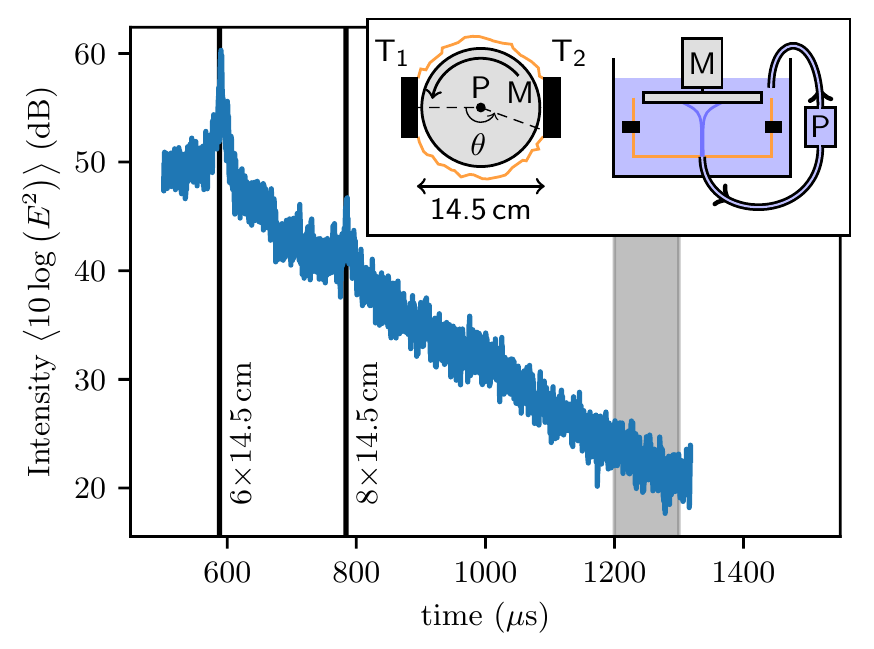}
\caption{Intensity of the acoustic wave emitted and received by $\mathsf{T_1}$ as a function of time. 
Two specular reflections are visible. The black vertical lines correspond to the propagation times of the wave for the distances indicated.
The gray domain corresponds to a typical time window over which the CBC is measured.
Inset: The experimental setup.}
\label{fig:IofT}
\end{figure}

The experimental setup is shown in the inset of Fig.~\ref{fig:IofT}.
A cylindrical metallic cavity (diameter, 15\,cm; height, 10\,cm) is filled with water.
Two coplanar linear arrays $\mathsf{T_1}$ and $\mathsf{T_2}$ of $N=64$ ultrasonic transducers centered at $f=6$\,MHz are placed face to face at the half-height of the cavity, and are separated from each other by $2d_0=14.5$\,cm.
Each transducer of the array is $L\simeq 10$\,mm high and $\mathrm{d}x=0.75$\,mm $\simeq\ 3 \lambda$ wide, where $\lambda$ is the ultrasonic wavelength.
A motor (Fig.~\ref{fig:IofT} inset, $\mathsf{M}$) drives a circular plate (diameter, 14\,cm) that is positioned at the top of the cylinder, and a pump (Fig.~\ref{fig:IofT} inset, $\mathsf{P}$) draws water through a hole (diameter, 8 \,mm) in the center of the bottom surface of the cylinder. In combination with the rotating plate, this creates a single and stable vorticity filament with core radius $r_0\sim \lambda$~\cite{Roux1997}.
Note that if the pump is turned off, the upper rotating plate creates solid rotation in the cylindrical cavity with a typical dimension $R\gg \lambda$~\cite{Lehmkuhl1971,Tourin1997}. 
In the first step, an ultrasonic plane wave can be transmitted through the vorticity filament by array $\mathsf{T_1}$, and the phase shift of the direct incident plane wave induced by the flow can be measured on array $\mathsf{T_2}$.
As $\mathrm{d}x\ll L$, this makes the plane wave propagate as a collimated beam that is perpendicular to the vortex axis, with a relatively small extension along the vortex direction.
The experimental configuration is thus close to the two-dimensional (2D) situation classically investigated in theoretical papers.

In the literature, the use of transducer arrays has provided spatial and dynamical characterization of rotating flows of different sizes~\cite{Roux1997,Rosny2005}.
When the vortex size is large compared to the ultrasound wavelength, the phase shift is easily interpreted using geometrical acoustics, and it yields a direct measurement of the vortex circulation, size, and position.
On the other hand, a vorticity filament (i.e., where the core size is smaller or comparable to the ultrasonic wavelength) behaves as a point-like scatterer in two dimensions.
Sound scattering by a single vorticity filament has been intensively studied following the publication of the Lighthill classical theory of aerodynamic sound~\cite{Lighthill1952,Fetter1964,OShea1975,Fabrikant1982,Sakov1993,Manneville2001}.
Theoretical and analytical analyses have been carried out to account for both sound scattering by the vortex core and long-range refraction effects due to the vortex flow~\cite{Reinschke1997,Ford1999}.

In 2D calculations, the scattered pressure field $\psi$ is analogous to the classical quantum mechanics problem of a beam of charged particles incident on a magnetic field tube, a problem known as the Aharonov-Bohm effect~\cite{Aharonov1959,Olariu1985}.
This formal analogy with quantum mechanics was first introduced by \citet{Berry1980}, who studied experimentally the scattering of surface waves by a bathtub vortex~\cite{Vivanco1999}.
The signature of the ultrasound scattering by a single vortex was experimentally observed by \citet{Roux1997}, who reported that the analytical calculations based on the quantum analogy also hold for acoustics. 

\begin{figure}
\includegraphics{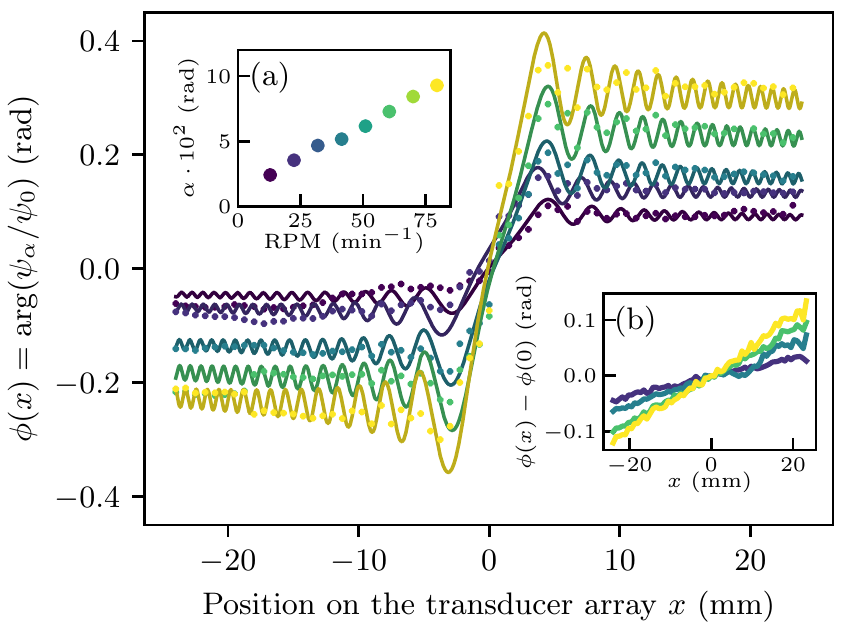}
\caption{Phase shift for the transmitted signal between $\mathsf{T_1}$ and $\mathsf{T_2}$ induced by a vorticity filament measured as a function of the rotation speed of the motor (color coded) at a given pump flow (0.06\,L$\cdot$s$^{-1}$).
The measurements are fitted to Eq.~(\ref{eq:vorticityPhase}) to extract the vorticity parameter $\alpha$ of the vortex (solid lines).
Inset (a): The fitted vorticity parameter $\alpha$ for the data plotted in the main figure as a function of the rotation speed of the upper circular plate.
Inset (b): Phase shift for the transmitted signal between $\mathsf{T_1}$ and $\mathsf{T_2}$ induced by a solid rotation (i.e., pump turned off).}
\label{fig:vortex}
\end{figure}

In practice, the vorticity filament is located at the center of the cavity around $x=0$.
The phase shift due to the wave-vorticity interaction on the transmitted signal between $\mathsf{T_1}$ and $\mathsf{T_2}$ is shown in Fig.~\ref{fig:vortex} for each of the transducers of the array $\mathsf{T_2}$.
\citet{Roux1997} showed that the phase jump and the phase oscillations are well described by the wave function predictions of the quantum Aharonov-Bohm effect~\cite{Olariu1985}
\begin{align}
\psi_\alpha(\rho,\theta) &= \sum_{m=-\infty}^\infty \exp\left(-\ii \frac{\pi}{2}\left|m-\alpha\right|\right)J_{\left|m-\alpha\right|} \left(k \rho\right)\e{\ii m \theta},
\label{eq:vorticityPhase}
\end{align}
where the origin of the polar coordinates $(\rho,\theta)$ is the vorticity filament, the angle $\theta$ is counted from the backscattering direction (see setup scheme in Fig.~\ref{fig:IofT}), and $k=2\pi/\lambda$ is the wave number.

The acoustic parameter $\alpha$ can be deduced from the quantum mechanics analogy $\alpha=\Gamma/\lambda c$, $c$ is the sound speed, and $\Gamma$ is the flow circulation. The fitting procedure of the curves is described in the Supplemental Material~\cite{SciPy,*Amos1986} and is shown in Fig.~\ref{fig:vortex}. The extracted $\alpha$ are plotted as a function of the angular velocity of the upper circular plate in Fig.~\ref{fig:vortex}(a).
For comparison, we measured the same quantity when the pump was turned off, i.e., for solid rotation of the water: linear dependency of the phase was measured along the array [see Fig.~\ref{fig:vortex}(b)], as expected by the radial velocity profile in the cavity for a solid rotation core larger than the array size. 

To measure the coherent backscattering effects in the cavity, a plane wave is now emitted from array $\mathsf{T_1}$, and the strongly reverberated and backscattered wave field is recorded on the same array. To increase the random scattering properties of the cylindrical cavity in the 2D propagation plane, a rough stainless steel sheet covers the inner boundary of the cylinder.
The role of the rough metallic sheet is to scatter the incident ultrasonic wave field and diminish the amplitude of the specular reflections, to obtain a quasi-2D cavity with rough boundaries. The intensity of a typical backscattered signal is shown in Fig.~\ref{fig:IofT}.
To further avoid the contributions of specular reflections (i.e., the peaks in Fig.~\ref{fig:IofT}), we choose to record the signal over $\delta t=0.1$\,ms after a travel time chosen between 1.2 and 1.8\,ms (e.g., the gray region in Fig.~\ref{fig:IofT} corresponds to such a typical time window). Following a method introduced by \citet{Aubry2007}, we then perform plane wave beam forming to recover the ultrasonic beams received in the directions around the incident plane-wave direction, with the power then calculated for the chosen time window.
Compared to earlier point-like measurements of the CBC~\cite{Tourin1997,Rosny2005} from an ultrasonic linear array, beam forming improves by a factor of 2 with the present array geometry both the signal-to-noise ratio of strongly dispersed signals and the angular resolution of the intensity distributions.

\begin{figure}
\includegraphics[width=\columnwidth]{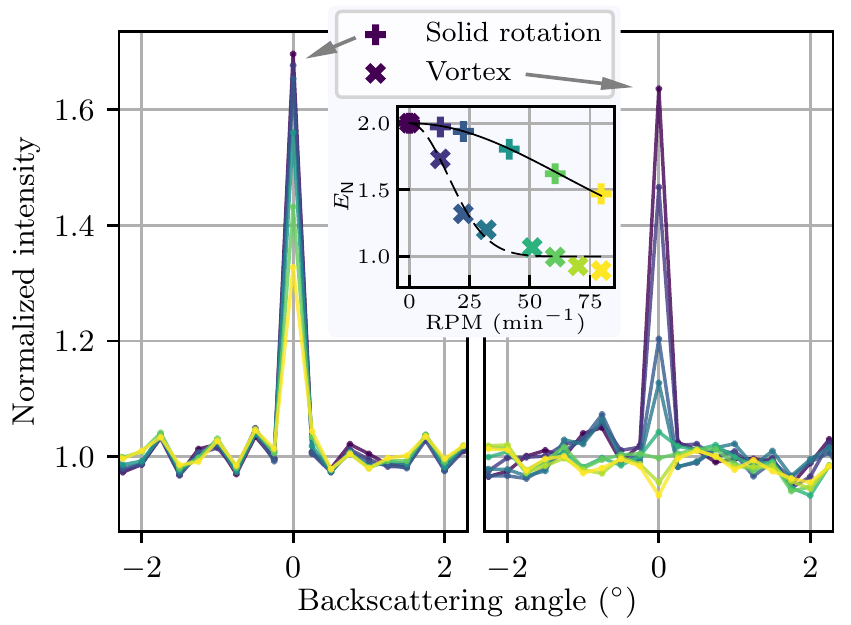}
\caption{Coherent backscattering cones measured with solid rotation (left, no pump), and measured with a vorticity filament (right, pump imposing a flow of 45\,mL$\cdot$s$^{-1}$).
Similar rotation speeds (color coded) were used without and with the pump.
The darkest curves on both sides show the same measurements realized without any flow.
Inset: Normalized enhancement factor $E_\mathsf{N}$ of the backscattering cones measured with solid rotation ($+$) and with a vorticity filament ($\times$), as a function of the rotation speed of the upper plate.}
\label{fig:CBC}
\end{figure}

To measure the CBC, there is the need to average over the disorder.
In optics, this is achieved automatically in colloidal dispersions, and dry samples need to be moved to average over the speckles.
Here, the cylindrical cavity has a fixed disorder that is given by the roughness of the metallic walls, and the positions of the transducers are fixed. Averaging over disorder can nevertheless be achieved by steering incident plane waves in different directions (typically every 0.5$^\circ$ from $-10^\circ$ to $10^\circ$), and by beam forming the received wave field around each incident angle. The intensity averaged over the different incident directions $I(\theta)$ is then normalized outside of the main peak by its average value for $1^\circ < \left|\theta\right|< 10^\circ$.
Such normalized intensities are shown in Fig.~\ref{fig:CBC}.
Note that the CBCs are measured here in a cavity with a typical width for the peak of $\lambda/\left(N\mathrm{d}x\right)$ (near field cone~\cite{Weaver1994,Rosny2000}), which is close to the angular resolution of the ultrasonic array.

The darkest CBC in both panels of Fig.~\ref{fig:CBC} are measured without any flow in the cavity. In this case, the wave propagation obeys the reciprocity, and the enhancement factor (i.e., the value of the normalized intensity at 0$^\circ$) is expected to be 2. However, this is not the case here, where the enhancement factor is about 1.7, which is explained by the diffraction-limited 3$\lambda$ width of each transducer element.
In the following, we define the normalized enhancement factor as $E_\mathsf{N}=\frac{E-1}{E_\mathrm{max}-1}+1$, $E_\mathrm{max}\simeq 1.7$, which is the value of $E$ without any flow.

The other CBCs of Fig.~\ref{fig:CBC} were measured with either solid rotation (left panel: upper plate turning, pump off) or a vorticity filament (right panel: upper plate turning, pump on).
The colors encode the rotation speed of the upper plate, and correspond to the colors of the points in the inset.
In both cases, the normalized enhancement factor decreases as a function of the rotation speed, as a signature of broken reciprocity~\cite{Erbacher1993,*Schertel2017}.
However, in the inset of Fig.~\ref{fig:CBC}, the normalized enhancement factor decreases more rapidly when there is a vortex filament in the cavity ($\times$), with respect to the case of solid rotation ($+$) for the same finite rotation speed. An interesting feature is the presence of a dip instead of a peak for the CBC at large rotation speed for the vorticity filament. This will be explained later in the Letter. 

The CBC shape depends on the path distributions inside the reverberating cavity.
In the case of solid rotation ($+$), the radial velocity profile is $\vec{v} = \Omega\, r\vec{u_\theta}$, and the dephasing between the counterpropagating paths is
\begin{align}
    \Delta\varphi=\frac{4\pi}{\lambda c}\oint \vec{v}\cdot\vec{\mathrm{d}\ell} = \frac{4\pi}{\lambda c} A\Omega,
    \label{eq:phaseSolidRotation}
\end{align}
where ${\Omega}$ is the vorticity field and $A$ is the area enclosed by the conjugated paths. The phase shift $\Delta\varphi$ is then path dependent.
Through calculation of the statistical properties of $A$ for randomly scattered paths in a cavity, \citet{Rosny2005} provided an expression for the enhancement factor as a function of the solid rotation,
\begin{align}
E(\Omega)\propto \exp{\left[-\frac{\pi^3R^3t}{\lambda^2 c}\Omega^2\right]},
\label{eq:rotation}
\end{align}
where $R$ is the radius of the solid rotation flow, $t$ is the travel time at which the measurements were carried out, and $\Omega$ is the vorticity of the flow.
For the present geometry, \citet{Lehmkuhl1971} experimentally measured the relation between $\Omega$ and the rotation frequency $f_0$ of the upper circular plate, as $\Omega\simeq 0.3 \cdot f_0$.
In the case of solid rotation, we fit the normalized enhancement factors $E_\mathsf{N}$ as a function of $f_0$ by a Gaussian curve [Eq.~(\ref{eq:rotation}), Fig.~\ref{fig:CBC} inset, solid line].
The only fit parameter---$R=4$\,cm---has an acceptable value in the case of solid rotation in terms of boundary conditions associated with the cavity size and the flow measurements shown in Fig.~\ref{fig:vortex}(b).
However, if we now perform the same Gaussian fit on the $E_\mathsf{N}$ measurements with a vorticity filament ($\times$, dashed curve in Fig.~\ref{fig:CBC}), the fit parameter---$R=10$\,cm---is not compatible with the vortex core size (comparable to the ultrasonic wavelength). 
This indicates that the different nature of both flows must fundamentally change their phase shift characteristics for strongly reverberated acoustic paths in the cavity.
In the following, we highlight the fundamental difference between solid rotation and vorticity filament flows, from the CBC point of view.

For a vorticity filament, the radial velocity profile is $\vec{v} = \frac{\Gamma}{2\pi r}\vec{u_\theta}$, which leads to a phase shift between counterpropagating paths [Eq.~(2)]
\begin{align}
    \Delta\varphi=\frac{4\pi}{\lambda c}n\Gamma= 4\pi\alpha\cdot n.
    \label{eq:phaseVortex}
\end{align}
In our experimental configuration, the Aharanov-Bohm parameter $\alpha$ results from the combination of the rotation speed of the upper plate, the flow imposed by the pump, and the frequency of the acoustic wave [Fig.~\ref{fig:vortex}(a)].
The striking result here is the topological stability of the phase difference: $\Delta\varphi$ depends solely on the \emph{discrete} number of loops~$n$ around the vortex core, independent of the path trajectory, in agreement with the Aharanov-Bohm prediction.
The phase shift acquired between two counterpropagating paths in the cavity in the presence of a vorticity filament [Eq.~(\ref{eq:phaseVortex})] imposes the normalized intensity of the interference between these in the backscattering direction as $1+\cos \Delta\varphi$.

Defining $p_n$ as the probability distribution of the number of loops $n$ around a given point in the cavity for closed paths---which is independent of the presence of the vorticity filament if the contribution of the field scattered by the vortex core is ignored---the enhancement factor $E$ is then expected to be
\begin{align}
    E(\alpha) &=\sum_{n\in \mathbb{N}} p_n \left[1+\cos(4\pi n\alpha)\right]
     \label{eq:EF}
    \end{align}
We assume a Poisson distribution for $p_n$ in agreement with the fact that loops occurring in a fixed time interval (corresponding to closed trajectories with a given path length) are rare events with an average probability that solely depends on the rough cavity geometry.
({This assumption is supported by numerical simulations of the number of loops done by closed paths in a 2D reverberating rough cavity, see Supplemental Material.})
We therefore write $p_n = \frac{\beta^n e^{-\beta}}{n!}$, where  $\beta = \left<n\right>$ is the average number of loops $n$ within a given travel time interval. It follows
    \begin{align}
   E(\alpha) &= 1+\Re\left\{\exp\left[\beta (e^{4i\pi \alpha}-1)\right]\right\}.
    \label{eq:EFbis}
\end{align}

\begin{figure}
\includegraphics{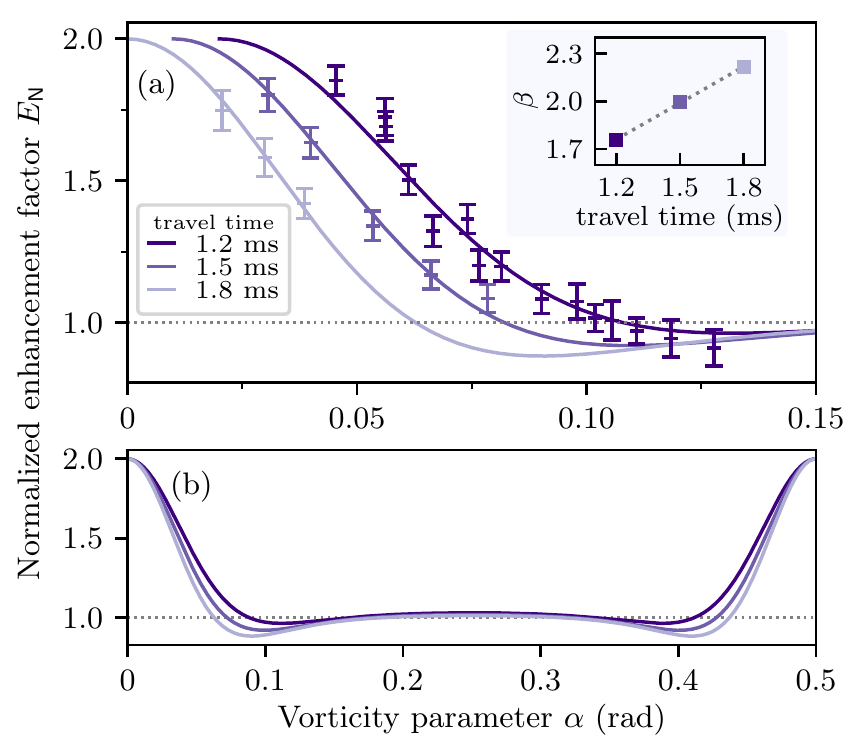}
\caption{(a) Enhancement factor of the coherent backscattering cone as a function of the vorticity parameter $\alpha$ for different travel times.
Each color corresponds to a travel time, and the corresponding theoretical curve is plotted in the same color.
For the sake of clarity, the three plots have been translated by $\alpha$=$0.01$.
Inset: Plot of the average loop number $\beta$ with respect to travel time.
(b) Predictions of the theory on a larger vorticity parameter range exhibiting a rebirth of the CBC for $\alpha=0.5$.}
\label{fig:EF}
\end{figure}

Figure~\ref{fig:EF}(a) shows the values of the normalized enhancement factors $E_\mathsf{N}$ as a function of $\alpha$ for different travel times, corresponding to different path lengths inside the cavity. Despite experimental dispersion explained by the uncertainty on the values of $E_\mathrm{max}$ used to compute $E_\mathsf{N}$, the measurements are in good agreement with theoretical prediction of Eq.~(\ref{eq:EFbis}), from which the average loop number can be extracted. As expected, $\beta$ linearly increases with travel time [inset in Fig.~\ref{fig:EF}(a)]. 

The model confirms the presence of a cone dip $E(\alpha)<1$ at intermediate values of $\alpha$ (see Fig.~\ref{fig:CBC}, right panel, yellow curve). The model also predicts that $E(\alpha)$ is $0.5$ periodic as shown in Fig.~\ref{fig:EF}(b).
In particular, the value $E(\alpha=0.5)=2$ corresponds to a surprising rebirth of the CBC in the absence of reciprocity which is a major difference compared to solid rotation, where $E$ strictly decreases to 1 for increasing rotation speed~\cite{Rosny2005}, and compared to the effects of Faraday rotation in optics~\cite{Erbacher1993,*Schertel2017}, where $E$ also irremediably decreases to 1 when the external magnetic field is increased.

However, can the ``rebirth'' of the CBC be obtained with realistic experimental configurations?
To answer this question, we assume a vortex core size $r_0 \sim \lambda$ and a maximum flow velocity $U_\mathrm{max}$, which leads to $\Gamma=2\pi r_0 U_\mathrm{max}$, and thus $\alpha = 2\pi\frac{ r_0 }{\lambda}\ \frac{U_\mathrm{max}}{ c}\sim 2\pi \mathrm{Ma}$, where $\mathrm{Ma}=U_\mathrm{max}/c$ is the flow Mach number.
Values of the vorticity parameter $\alpha = 0.5 $ then induce $\mathrm{Ma} \sim 0.1$, which is impossible to achieve in water because of the appearance of cavitation bubbles in the vortex core, which slow down the flow velocity.
However, $\mathrm{Ma} \sim 0.1$ might be reachable with one single vorticity filament in a medium with lower sound velocities, such as air~\cite{Pinton2005}, second sound in superfluid helium~\cite{Donnelly2009}, or in ultracold atomic gases~\cite{Sidorenkov2013}.

To conclude, we have shown in this Letter how a vorticity filament behaves as a topological anomaly for wave propagation that ``counts'' the number of loops made by long acoustic paths around it. Consequences of the topological stability of the phase difference between counterpropagating paths are (1) the coherent backscattering dip experimentally observed for $\alpha\gtrsim 0.1 $  and (2) the prediction of the rebirth of the coherent backscattering cone for $\alpha = 0.5 $ under strongly reciprocity-breaking conditions.

\begin{acknowledgments}
The authors thank Georg Maret, and G.J.A. acknowledges support from the Zukunftskolleg (Universität Konstanz) for a Mentorship grant. The authors wish to thank Frédéric Faure, Julien de Rosny, Bart van Tiggelen and Arnaud Tourin for fruitful discussions.
The authors further acknowledge the anonymous referees, especially for the suggestion of using the Poisson distribution.
\end{acknowledgments}

\appendix
\setcounter{figure}{0}
\renewcommand{\thefigure}{S\arabic{figure}}
\setcounter{equation}{0}
\renewcommand{\theequation}{S\arabic{equation}}

\section{Supplemental Material}

The Supplemental Material contains information on the fitting procedure of the phase shifts of Fig.~\ref{fig:vortex}, and on the numerical simulations justifying the use of the Poisson distribution in Eq.~(\ref{eq:EFbis}).

\subsection{Phase shift fits of Fig.~\ref{fig:vortex}}

In order to fit the phase shifts of Fig.~\ref{fig:vortex}, we first convert the positions $x$ on the transducer array into the polar coordinates $(r,\theta)$ used in Eq.~(\ref{eq:vorticityPhase}), as defined in the inset of Fig.~\ref{fig:IofT},
\begin{align}
    \theta(x) &= \pi - \arctan\frac{x}{d_0},\\
    \rho(x) &= \frac{x}{\sin \left(\pi - \theta(x)\right)}.
\end{align}

The experimental data are then fitted by the following formula
\begin{align}
    \phi(x)=\psi_\alpha\left(\rho(x+x_0),\theta(x+x_0)\right) + \phi_0
\end{align}
where $\psi_\alpha$ is defined by Eq.~(\ref{eq:vorticityPhase}) with the Aharonov-Bohm parameter $\alpha$, and two other fit parameters, $x_0$ that takes into account that the vorticity filament is not directly facing the center of the transducer array, and $\phi_0$ that account for a small phase shift adjustment between the two arrays.
In Eq.~(\ref{eq:vorticityPhase}), $k=\frac{2\pi f}{c}$ with $c=1500$\,m/s in water, and the sum was performed between $m=-2000$ and $m=2000$.
The Bessel function of the first kind $J_{\left|m-\alpha\right|}$ in Eq.~(\ref{eq:vorticityPhase}) was calculated using the SciPy wrapper~\cite{SciPy} of the AMOS \texttt{zbesj} routine~\cite{Amos1986}.
On the curves of Fig.~\ref{fig:vortex}, the two other fit parameters ranges are $-0.8\,\mathrm{mm}<x_0<0.1\,\mathrm{mm}$ and $0.01\,\mathrm{rad}<\phi_0<0.04\,\mathrm{rad}$.
Let us mention that fixing these two additional parameters to zero do not change quantitatively the values of the extracted $\alpha$. Moreover, we also checked that taking into account the scattering by a finite size vorticity filament [Eq.~(2.126) of ref.~\cite{Olariu1985}] provide very similar $\alpha$ values, and very small vorticity filament core size compared to the ultrasonic wavelength.

%

\subsection{Numerical simulations of the number of loops in the cavity}

\begin{figure}
\begin{minipage}{0.48\textwidth}
    \centering
    \includegraphics{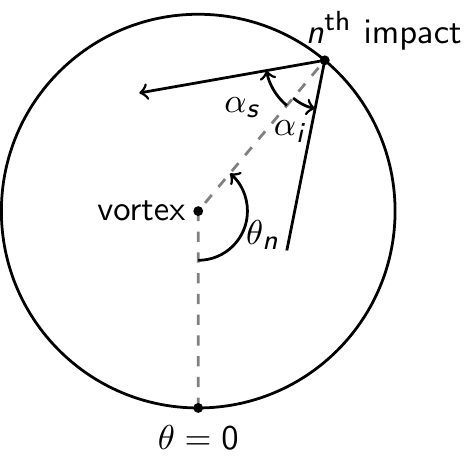}
    \caption{Geometry of the numerical simulations}
    \label{fig:simulations}
\end{minipage}
\begin{minipage}{0.48\textwidth}
    \centering
    \includegraphics{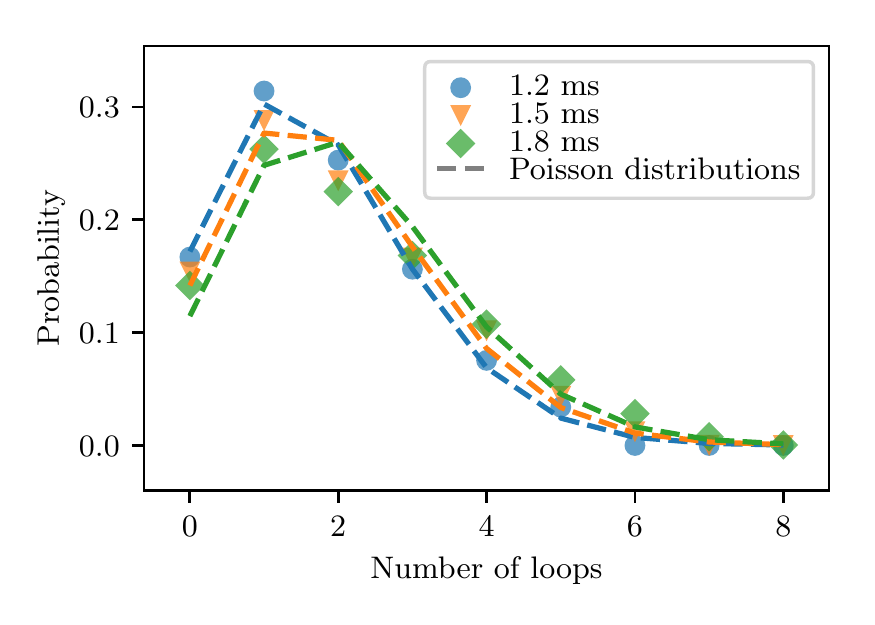}
    \caption{Statistics of the number of loops done by the random walks described in the text for different random walk lengths (symbols) and the corresponding Poisson distributions (dashed lines, the colors encode the path lengths).}
    \label{fig:poisson}
\end{minipage}
\end{figure}

In order to get an estimation of the number of loops done by each closed paths in the cavity, we perform an angular random walk in polar coordinates.
The geometry is shown in Fig.~\ref{fig:simulations}.
The random walks starts at the bottom point ($\theta=0$), with a random angle, and travels on straight lines between successive scattering events.
A scattering event happens each time the random walk crosses the circle.
Let us call $\alpha_\mathrm{i}$ the angle with respect to the normal at which the path impacts the circle, the path then continues with an angle $\theta_\mathrm{s}$ which is chosen on a normal law centered on $\alpha_\mathrm{s}=\alpha_\mathrm{i}$ (specular reflection) with a certain width (standard deviation $\sigma$) accounting for the roughness of the surface.
We continue the random walk until the total travel time of the walk is equal to the measured ones (1.2\,ms, 1.5\,ms or 1.8\,ms; radius of the cavity, 7.25\,cm; sound velocity, 1500\,m/s).
We are only interested in the statistics of the backscattered paths, which mean that we only keep the paths which end at $\theta=0$ after the total travel time.
During the random walks, we log the number of loops done by the path around the vortex.
Fig.~\ref{fig:poisson} shows an example of the statistics of the number of loops done by 50,000 random walks ending at $\theta=0$, when $\sigma=40^\circ$.
The dashed lines correspond to the Poisson distributions calculated using the average number of loops extracted from the statistics.
The good agreement between the simulated statistics and the corresponding Poisson distributions justifies Eq.~(\ref{eq:EFbis}).

\bibliography{acousticBiblio}

\end{document}